%% file: main.tex
\documentclass{article}

\usepackage[noblocks]{authblk}
\usepackage{cite}
\usepackage[pdftex]{graphicx}
\graphicspath{{img/}}
\DeclareGraphicsExtensions{.png}
\usepackage{siunitx}
\usepackage{url}

\begin{document}

\title{The Unreasonable Effectiveness of Address Clustering}

\author{Martin Harrigan\thanks{Email: martinharrigan@gmail.com}}
\affil{Waterford Institute of Technology}

\author{Christoph Fretter}
\affil{Elliptic Enterprises Limited, London}

\date{}

\maketitle

\begin{abstract}
\input{tex/abstract}
\end{abstract}

\input{tex/intro}

\input{tex/related}

\input{tex/counts}

\input{tex/sizes}

\input{tex/formation}

\input{tex/conc}

\bibliographystyle{plain}

\bibliography{bib/main}

\end{document}

%% file: tex/abstract.tex
Address clustering tries to construct the one-to-many mapping from
entities to addresses in the Bitcoin system. Simple heuristics based
on the micro-structure of transactions have proved very effective in
practice. In this paper we describe the primary reasons behind this
effectiveness: address reuse, avoidable merging, super-clusters with
high centrality, and the incremental growth of address clusters. We
quantify their impact during Bitcoin's first seven years of existence.

%% file: tex/intro.tex
\section{Introduction}
\label{sec:introduction}

Bitcoin is a double-edged sword for financial privacy. It allows
anyone to conduct financial transactions with anyone else in the world
without having to divulge identifying information to
intermediaries. However, it requires those transactions to be
broadcast to the world. The contents of the transactions, their
relationship with other transactions, and the very act of broadcasting
the transactions themselves may unintentionally disclose information
about the transactors to interested third parties. In fact, many
interested third parties systematically gather and analyze this
information for reasons such as market research, competitor analysis,
compliance, and law enforcement.

Address clustering is a cornerstone of this analysis. It partitions
the set of addresses observed in Bitcoin transactions into maximal
subsets of addresses that are likely controlled by the same
entity. Each subset in the partition is an address cluster. When
combined with address tagging (associating real-world identities with
addresses) and graph analysis, it is an effective means of analysing
Bitcoin activity at both the micro- and macro-levels, see, e.g.,
\cite{reid-harrigan-13,meiklejohn-et-al-13,androulaki-et-al-13,ober-et-al-13,ortega-13,spagnuolo-et-al-13,fleder-et-al-15,lischke-fabian-16}.
Experimental analysis has shown that a single
heuristic (the multi-input heuristic) can identify more than
\num{69}\% of the addresses in the wallets stored by lightweight
clients.

\begin{figure*}
  \includegraphics[width=\textwidth]{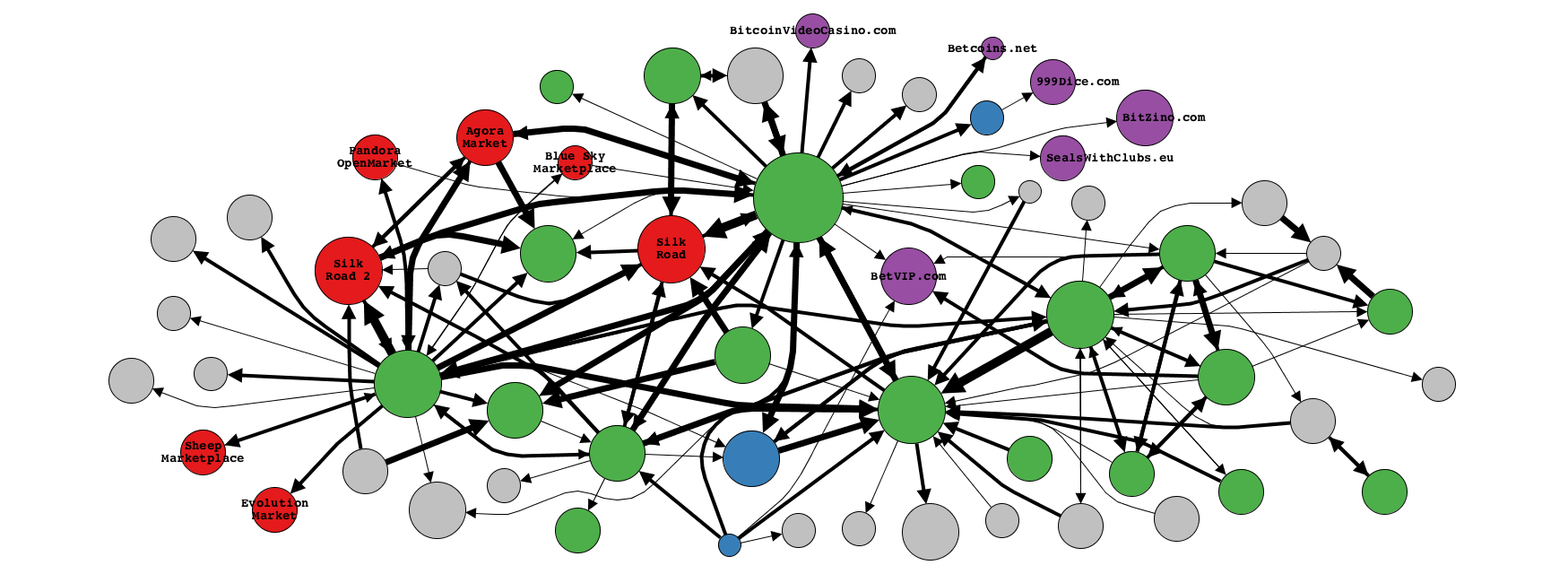}
  \caption{A graphical summary of the most significant flows of
    bitcoin between the largest address clusters during Bitcoin's
    first five years in existence. The vertices correspond to address
    clusters: red vertices are darknet markets; purple vertices are
    gambling services; green vertices are exchanges and blue vertices
    are mining pools. The gray vertices are not immediately
    identifiable using publicly available information.}
  \label{fig:top-address-clusters}
\end{figure*}

As a token of its effectiveness, consider
Fig~\ref{fig:top-address-clusters}. This is a graphical summary of
the most significant flows of bitcoin between the largest address
clusters during Bitcoin's first five years in existence. Using
publicly available information, we can identify all but the gray
vertices: the red vertices are darknet markets; the purple vertices
are gambling services; the green vertices are exchanges and the blue
vertices are mining pools. The labels for the exchanges and mining
pools, although known, are omitted to avoid indiscriminately linking
their identities to darknet markets without fully presenting the
methodology behind this summary and the definitions for ``most
significant flows" and ``largest address clusters". However, it is
based on the methodologies presented in the papers above and relies on
address clustering at its core.

This paper considers the reasons behind the effectiveness of address
clustering using the multi-input heuristic~\cite{nakamoto-08}. This
heuristic assumes that the addresses in transaction outputs redeemed
in a multi-input transaction were controlled by the same
entity. Although not true in the general case, it is a useful
heuristic in practice. In Sect.~\ref{sec:related-work} we briefly list
some related work. In Sections~{\ref{sec:counts} and \ref{sec:sizes}}
we study address cluster counts and sizes. We quantify the levels of
address reuse and cluster merging. We observe ``super-clusters'' and
analyze their centrality. We study the formation and structure of
address clusters in Sect.~\ref{sec:formation}. We conclude with some
future work in Sect.~\ref{sec:conclusion}.

%% file: tex/related.tex
\section{Related Work}
\label{sec:related-work}

Address clusters are the fundamental building-blocks on which many
high-level blockchain analyses are performed. They can be constructed
using the multi-input heuristic as noted by Bitcoin's
creator~\cite{nakamoto-08}. Reid and Harrigan~\cite{reid-harrigan-13}
considered the impact of address clusters on anonymity. This approach
can be augmented with change
heuristics~\cite{androulaki-et-al-13,meiklejohn-et-al-13,spagnuolo-et-al-13},
temporal behavior~\cite{ortega-13,monaco-15} and transaction
fingerprinting~\cite{fleder-et-al-15}. Although the analyses in the
present paper are based on the multi-input heuristic only, they can be
extended to any combination of heuristics.

Nick~\cite{nick-15} analyzed the performance of several clustering
heuristics by exploiting a vulnerability in connection Bloom filtering
used by lightweight clients. He found that the multi-input heuristic
can identify more than \num{69}\% of the addresses in the vulnerable
wallets.

Ober et al.~\cite{ober-et-al-13} studied the sizes and lifespans of
address clusters and showed that the sizes of the address clusters
follow a scale-free distribution. Lischke and
Fabian~\cite{lischke-fabian-16} showed that major darknet markets,
gambling services, exchanges and mining pools were major hubs in the
address cluster graph (similar to Fig.~\ref{fig:top-address-clusters}
but not limited to the largest address clusters) during Bitcoin's
first four years of existence.

Maxwell described CoinJoin~\cite{maxwell-13}, a protocol for trustless
centralized Bitcoin mixing. This causes the multi-input heuristic to
produce false positives. CoinJoin is a centralized mixing protocol; it
requires a third-party or CoinJoin server to operate. Other protocols
in this category include Mixcoin~\cite{bonneau-et-al-14} and
Blindcoin~\cite{valenta-rowan-15}. Decentralized mixing protocols do
not require any third-party, trusted or trustless. Protocols in this
category include CoinSwap~\cite{maxwell-13-1},
CoinShuffle~\cite{ruffing-et-al-14} and
CoinParty~\cite{ziegeldorf-et-al-15}. Shentu and
Yu~\cite{shentu-yu-15} review several trustless Bitcoin protocols.

M{\"o}ser et al.~\cite{moeser-et-al-13,moeser-et-al-14} considered the
implications of blockchain analyses, including address clustering, for
anti-money laundering. Imwinkelreid~\cite{imwinkelried-et-al-15}
discussed its implications for digital forensics.

%% file: tex/counts.tex
\section{Counting Address Clusters}
\label{sec:counts}

The following analyses were performed when the block at the tip of the
Bitcoin blockchain was at height \num{396577} and the last eight
hexadecimal digits of the block hash were \texttt{900a6f4c}.

\begin{figure}
  \includegraphics[width=\columnwidth]{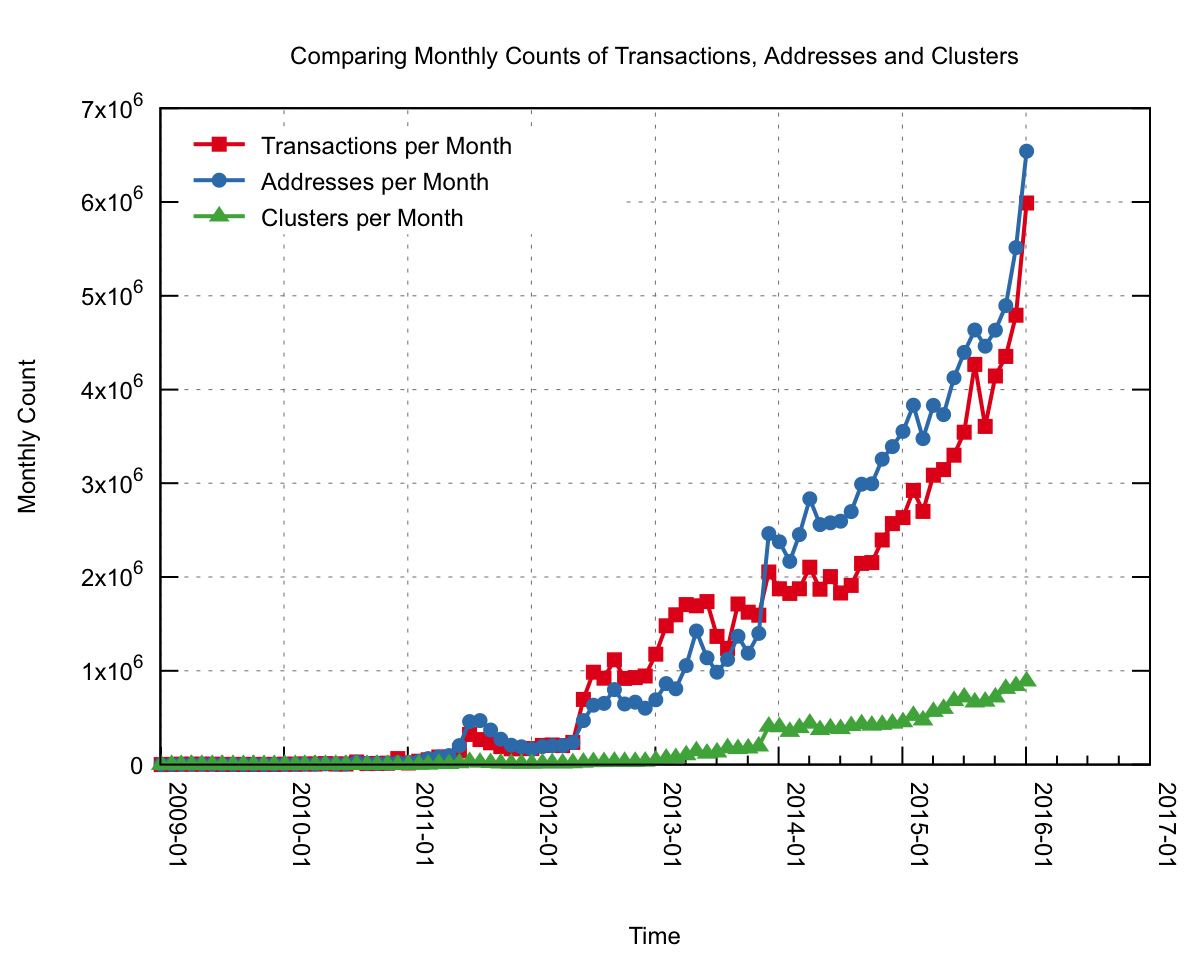}
  \caption{A plot of the monthly counts of transactions (red line),
    new addresses (blue line) and address clusters (green line) with
    at least two addresses. For the past two years, the monthly number
    of new addresses is greater than the monthly number of
    transactions.}
  \label{fig:counts}
\end{figure}

Figure~\ref{fig:counts} compares the monthly counts of transactions
(red line) with the monthly counts of new addresses (blue line). The
number of new addresses has grown in line with the number of
transactions. The monthly counts of address clusters (green line) with
at least two addresses has grown at a much slower rate.

\begin{figure}
  \includegraphics[width=\columnwidth]{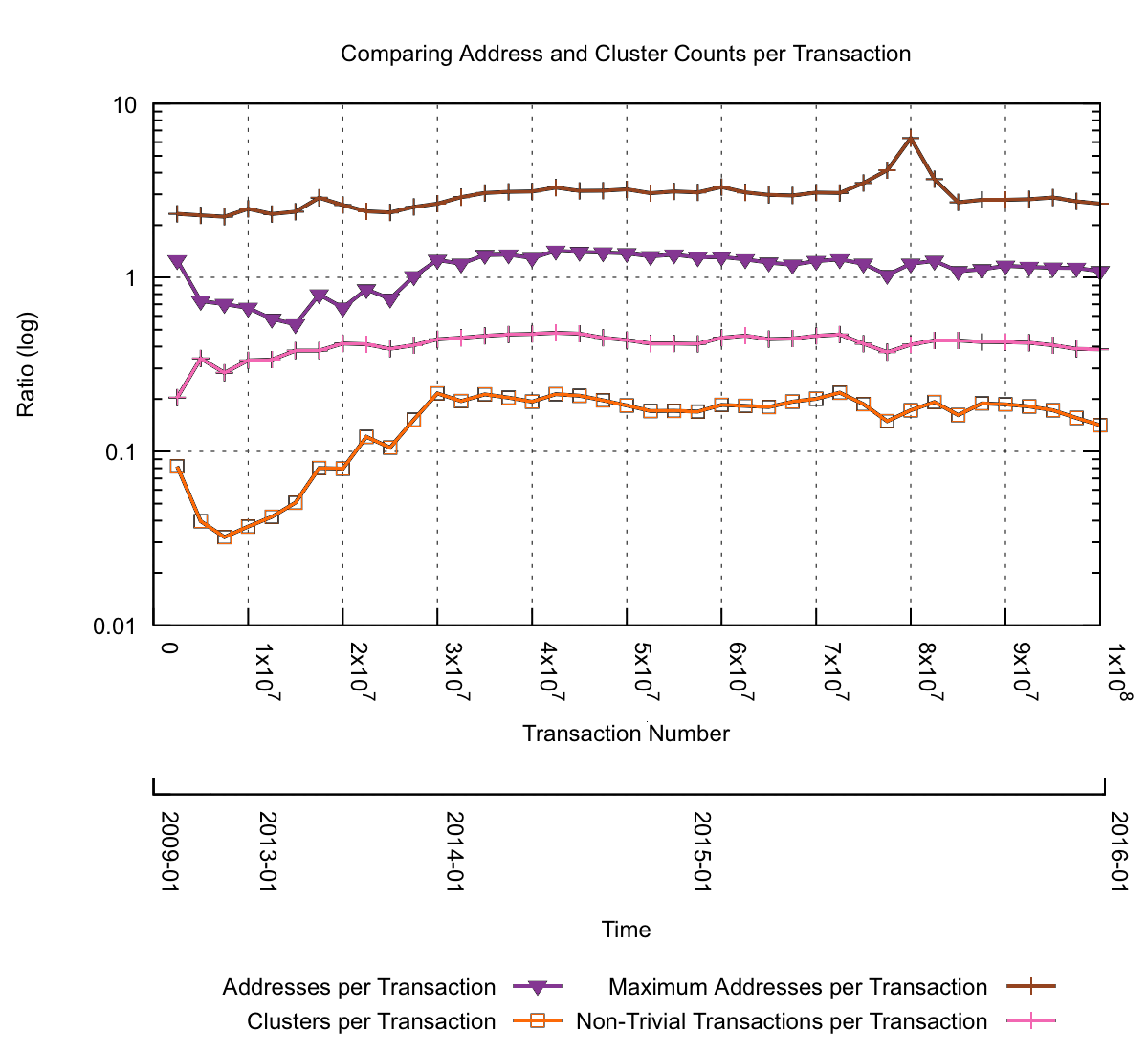}
  \caption{A plot of the ratios of new addresses per transaction
    (purple line) and newly merged address clusters per transaction
    (orange line). The maximum addresses per transaction (brown line)
    and non-trivial transactions per transaction (pink line) are
    respective upper bounds.}
  \label{fig:ratios-annotated}
\end{figure}

We consider the relationship between these counts in
Fig.~\ref{fig:ratios-annotated}. We plot the number of new addresses
observed per transaction (purple line) and the number of newly merged
address clusters created per transaction (orange line). To adjust for
the rapid growth in transactions in Bitcoin's recent history, we
replace the horizontal time axis with ordinal transaction numbers:
this compresses low-activity periods and expands high-activity
periods. We observe that both ratios have been relatively stable for
the past two years and that the former is an order of magnitude larger
than the latter.

Can we establish upper bounds for the two ratios?
Nakamoto~\cite{nakamoto-08} suggested that ``a new key pair should be
used for each transaction to keep them from being linked to a common
owner." This is from the perspective of the payee(s) only; if the
payer requires additional transaction outputs, say, for change, they
should also use a new address. For transaction outputs that contain
Pay-to-PubKey and Pay-to-PubKey-Hash scripts, the number of
transaction outputs per transaction is an upper bound for the number
of new addresses per transaction. This can be adjusted for transaction
outputs that contain \texttt{OP\_RETURN} scripts, multi-signature
scripts and Pay-to-Script-Hash scripts where the redemption script is
known (brown line). The gap between the brown and purple lines is a
measure of the level of address reuse; the wider the gap the greater
the level of address reuse.

Similarly, the fraction of transactions that spend at least two
transaction outputs assigned to different addresses (pink line) is an
upper bound for the number of newly merged address clusters per
transaction. We refer to these transactions as being non-trivial. If
every transaction output created a new address then every non-trivial
transaction would create a newly merged address cluster. The gap
between the pink and orange lines is a measure of the level of cluster
merging; the wider the gap the greater the level of cluster merging.
Even in the presence of address reuse, this gap could be narrowed
through the use of merge avoidance~\cite{ranvier-13,hearn-13}.

The existence of both gaps is significant. New key pairs are not being
generated for every transaction allowing the multi-input heuristic to
link addresses to a common owner. This is one reason that address
clustering is unreasonably effective. There is considerable reuse of
addresses and merging of address clusters. We will discuss a second
reason in the next section.

%% file: tex/sizes.tex
\section{Measuring Cluster Sizes}
\label{sec:sizes}

\begin{figure}
  \includegraphics[width=\columnwidth]{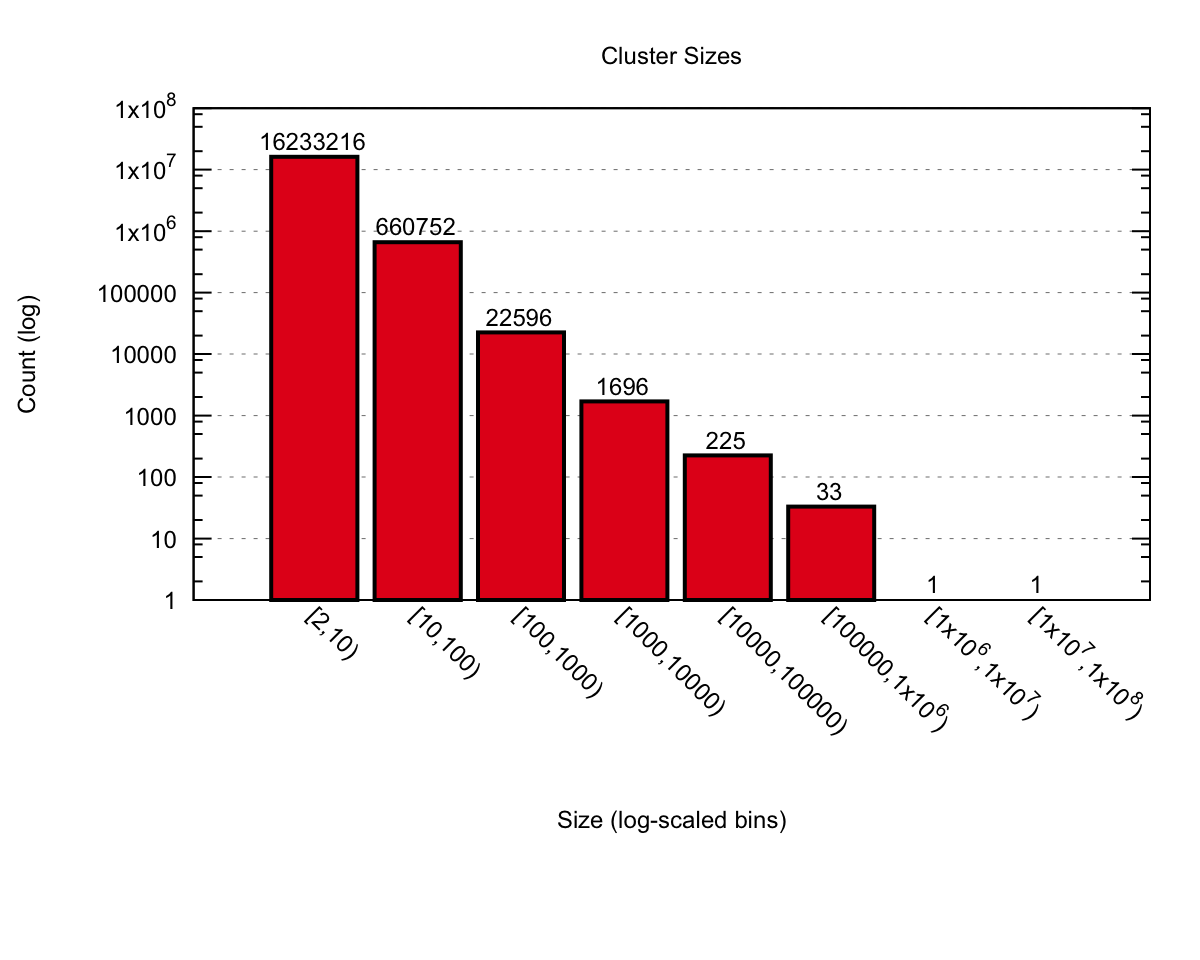}
  \caption{A histogram showing the number of address clusters with at
    least two addresses in each size range.}
  \label{fig:sizes}
\end{figure}

The address clusters with at least two addresses are binned by size in
Fig.~\ref{fig:sizes}. Both the horizontal and the vertical axes use
logarithmic scales. We observe the presence of ``super-clusters'':
there are \num{1955} address clusters with at least \num{1000}
addresses but less than \num{10} million addresses. They cover
\num{22}\% of all of the addresses represented in Fig.~\ref{fig:sizes}
and \num{16}\% of all of the addresses observed at the time of the
analysis.

We exclude the single address cluster with greater than \num{10}
million addresses. This address cluster originally belonged to the
Mt.\ Gox exchange that, for a time, allowed users to import
private-keys directly from their wallets. This feature causes the
multi-input heuristic to produce false positives and requires more
advanced heuristics to separate the Mt.\ Gox addresses. We will
discuss this issue in Sect.~\ref{sec:formation}.



The super-clusters are not only large in terms of the number of
addresses they contain, they are also hubs in terms of the number of
transactions they are involved in. At the time of the analysis, the
\num{107} million transactions created \num{319} million transaction
outputs and redeemed \num{285} million of those through transaction
inputs. Of those, the super-clusters were responsible for \num{72}
million or \num{23}\% of the transaction outputs and \num{51} million
or \num{18}\% of the transaction inputs. If we can link identities to
the super-clusters then we can identify at least one of the
transactors in a considerable number of transactions.

Lischke and Fabian~\cite{lischke-fabian-16} made a related
finding---they analyzed the degree centrality of the vertices in a
network similar to the one in Fig.~\ref{fig:top-address-clusters} but
not limited to the largest address clusters, and found that the
vertices representing the major darknet markets, gambling services,
exchanges and mining pools had the highest degree centralities.

The existence and centrality of super-clusters is another reason that
address clustering is unreasonably effective. Many of the major
services reuse addresses and generate super-clusters thereby
identifying much of their on-chain activity. Furthermore, this
identifies much of the activity between the service and their users:
deposits and withdrawals can be easily identified. This can be
exploited to produce ``wallet explorers'' such as
\url{WalletExplorer.com}. It also makes the services vulnerable to
re-identification attacks~\cite{meiklejohn-et-al-13}.

Major services can avoid creating super-clusters. For example,
Coinbase, the Bitcoin exchange and wallet provider, does not create a
super-cluster that identifies all activity between the service and
their users. They are notably absent from many high-level blockchain
analyses. This is not to say that they do not create any large
clusters. It simply means that the multi-input heuristic alone is
insufficient for identifying all of their on-chain activity.

%% file: tex/formation.tex
\section{Formation and Structure}
\label{sec:formation}

\begin{figure}
  \includegraphics[width=\columnwidth]{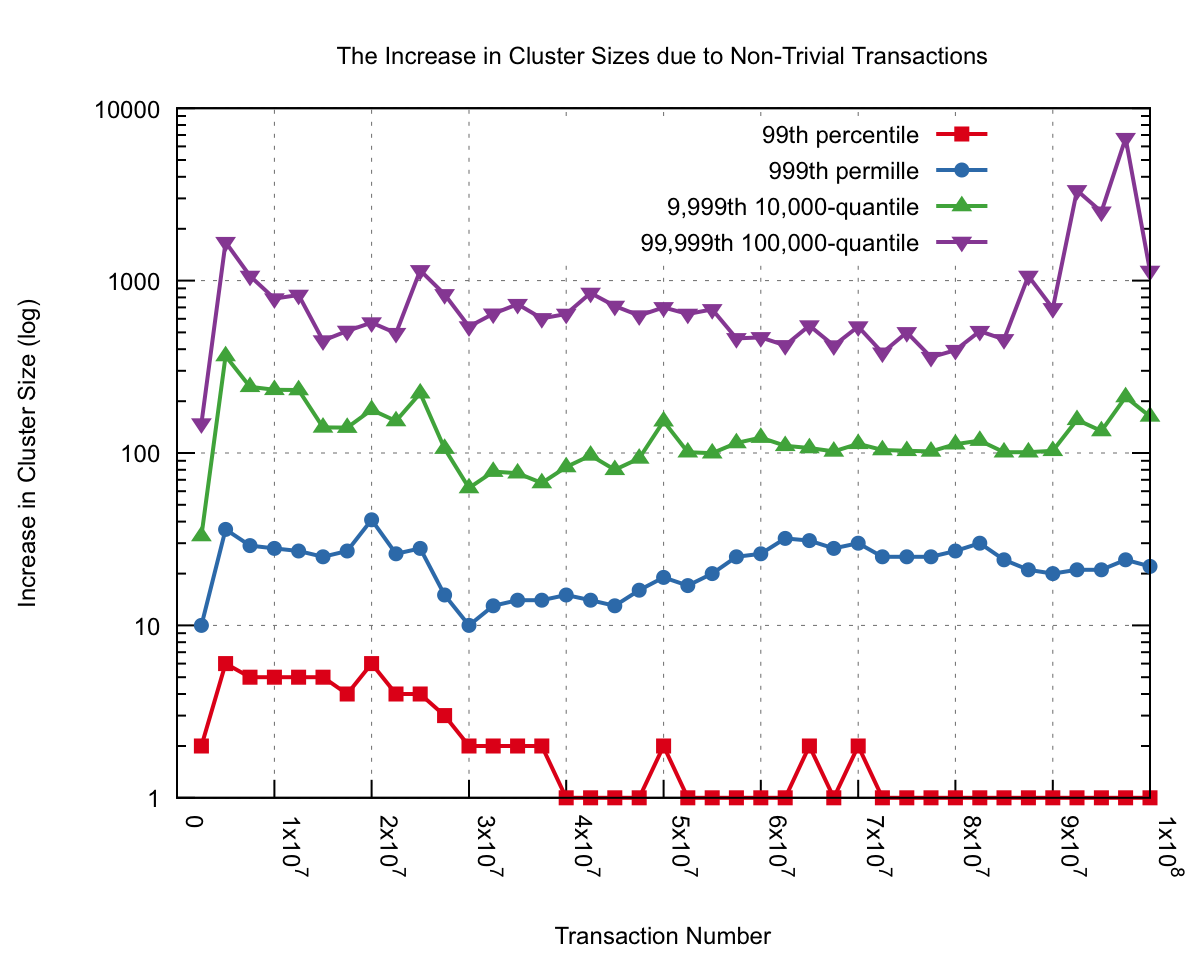}
  \caption{A plot of the \(q -1\)th \(q\)-quantiles for \(q = 100,
    1000, 10000, 100000\) of the distributions of the increases in
    cluster size due to merging for every \num{250000}
    transactions. The increases are heavily concentrated around median
    values of one. For the past \num{30} million transactions, the
    \num{99}th percentiles are also at one.}
  \label{fig:increases}
\end{figure}

The address clustering heuristics listed in
Sect.~\ref{sec:related-work} cause address clusters to merge. We are
not aware of any published heuristics that cause address clusters to
split, e.g.\ to counter the mixing protocols in
Sect.~\ref{sec:related-work} or to partition the Mt.\ Gox address
cluster in Sect.~\ref{sec:sizes}. When address clusters merge, we can
measure the increases in size of the newly merged cluster. For
example, suppose a transaction causes four address clusters of sizes
\num{1}, \num{1}, \num{2} and \num{10} to merge. This can be
represented by increases of \num{1}, \num{1} and \num{2}. Considering
the multi-input heuristic only, the distribution of these increases is
heavily concentrated around a median value of
one. Figure~\ref{fig:increases} plots the \num{99}th percentile,
\num{999}th permille, \num{9999}th \num{10000}-quantile and
\num{99999}th \num{100000}-quantile for every \num{250000}
transactions. We observe that large increases in address cluster sizes
are rare. The multi-input heuristic usually merges at most one large
address cluster with one or more small address clusters, but rarely
merges two or more large address clusters.

\begin{figure}
  \includegraphics[width=\columnwidth]{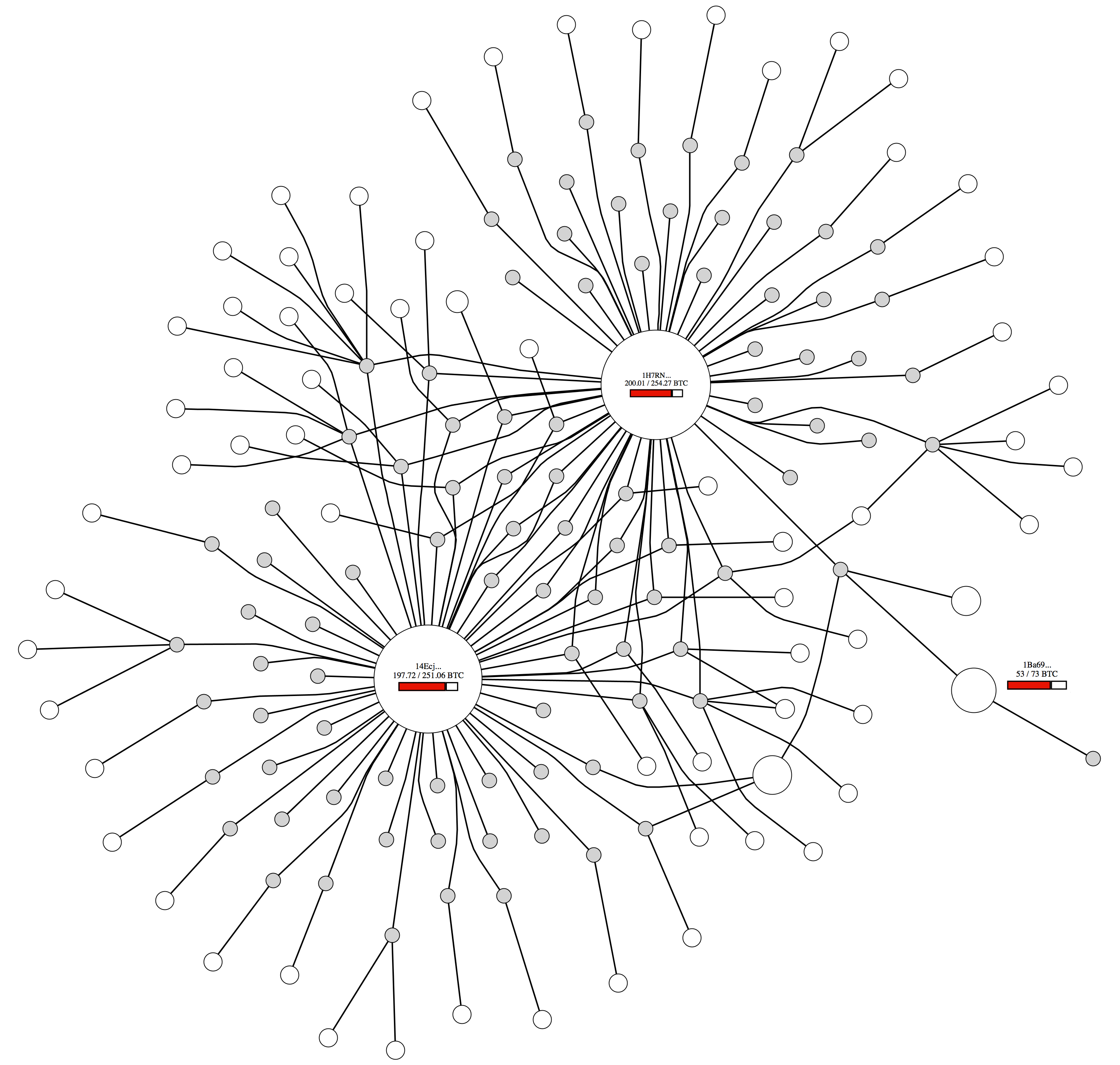}
  \caption{A bipartite graph representing the structure of an address
    cluster: white vertices are addresses; gray vertices are
    transactions; edges connect transaction vertices to address
    vertices when the corresponding transaction spends transaction
    outputs that were assigned to the corresponding addresses.}
  \label{fig:co-spents}
\end{figure}

This behaviour can be visualised as follows. Consider a bipartite
graph for each address cluster generated using the multi-input
heuristic where each vertex represents either an address (an address
vertex) or a transaction (a transaction vertex) and each edge between
an address vertex and a transaction vertex represents the transaction
spending a transaction output that was controlled by the
address. Figure~\ref{fig:co-spents} is the bipartite graph for a
typical address cluster\footnote{The address cluster contains the
  address \texttt{1H7RNFmAbtMgVJzK72hNFerGBfKuRekTMU}: it received
  bitcoins from a mining pool (DeepBit), sent bitcoins to exchanges
  (Mt.\ Gox and \url{bitcoin.de}) and purchased goods through a
  Bitcoin payment processor (BitPay).}. The white vertices are the
address vertices. The address vertices that correspond to addresses
with non-zero balances are annotated with their current and all-time
maximum balances. The majority of addresses have zero balances. The
gray vertices are the transactions---they connect together the
addresses to form the address cluster. Address vertices that are
connected through multiple independent paths have multiple independent
sets of transactions indicating that they are part of the same address
cluster.

This graph was formed by small address clusters (the singleton address
vertices along the periphery) merging with the large address cluster,
through transaction vertices that connected the singleton address
vertices to address vertices with non-zero balances. It is rare for
such a graph to form as two large disconnected components, each
containing at least one address vertex with a non-zero balance, and
then to merge into a single connected component.

Address clusters that form when two large address clusters merge can
be flagged as exhibiting unusual merging activity. This can be
extended to a heuristic for splitting address clusters that may not be
controlled by the same entity. For example, if we identify the
\num{0.01}\% of transactions that resulted in the largest increases in
cluster size during the lifetime of the Mt.\ Gox exchange (July 2010
to~February 2014), then the majority of those transactions spend
transaction outputs that were controlled by the Mt.\ Gox address
cluster. This is likely due to their private-key import feature.

The incremental growth of address clusters is beneficial for many
high-level blockchain analyses. The address clustering is relatively
stable over time. It is a rarity for two large address clusters to
merge, thereby drastically changing the results of an earlier
analysis. If fact, if two large address clusters do merge, it may
indicate that the multi-input heuristic has produced a false
positive. Furthermore, the address clustering is suitable for
real-time analyses. Small address clusters merge with large address
clusters early in their lifetime and the large address clusters are
more likely to have identifying information associated with them.

%% file: tex/conc.tex
\section{Conclusion and Future Work}
\label{sec:conclusion}

We have enumerated and analyzed the primary reasons behind the
effectiveness of address clustering using Bitcoin's blockchain. These
are the high-levels of address reuse and avoidable merging; the
existence of super-clusters with high centrality, and the incremental
growth of address clusters.

The results can inform and help blockchain analysts. For example, the
super-clusters are primary targets for re-identification attacks. The
technique at the end of Sect.~\ref{sec:formation} can flag address
clusters that may include addresses from more than one entity.

The opposing camp, those seeking to hinder blockchain analysis, can
also benefit from these results. For example, the adoption and impact
of privacy-enhancing techniques such as merge avoidance and Elliptic
Curve Diffie-Hellman-Merkle (ECDHM) address schemes, e.g.\ stealth
addresses~\cite{todd-14}, reusable payment codes (BIP47)~\cite{bip-47}
and out of band address exchange (BIP75)~\cite{bip-75}, can be
measured indirectly through the gap between the number of non-trivial
transactions and the number of address clusters created or merged per
transaction (see Sect.~\ref{sec:counts}).

Our future work revolves around the internal structure of address
clusters, {\`a} la the bipartite graph in
Fig.~\ref{fig:co-spents}. This representation shows the structure of
an address cluster beyond a simple set of addresses and may provide
further insight into its formation and behavior.

%% file: main.bbl
\begin{thebibliography}{10}

\bibitem{androulaki-et-al-13}
Elli Androulaki, Ghassan Karame, Marc Roeschlin, Tobias Scherer, and Srdjan
  Capkun.
\newblock {E}valuating {U}ser {P}rivacy in {B}itcoin.
\newblock In {\em Proceedings of the 17th International Conference on Financial
  Cryptography and Data Security, FC 2013, Okinawa, Japan, April 1--5, 2013},
  pages 34--51, 2013.

\bibitem{bonneau-et-al-14}
Joseph Bonneau, Arvind Narayanan, Andrew Miller, Jeremy Clark, Joshua Kroll,
  and Edward Felten.
\newblock {M}ixcoin: {A}nonymity for {B}itcoin with {A}ccountable {M}ixes.
\newblock In {\em Proceedings of the 18th International Conference on Financial
  Cryptography and Data Security, FC 2014, Christ Church, Barbados, March 3--7,
  2014}, pages 486--504, 2014.

\bibitem{fleder-et-al-15}
Michael Fleder, Michael~S. Kester, and Sudeep Pillai.
\newblock {B}itcoin {T}ransaction {G}raph {A}nalysis.
\newblock {\em CoRR}, abs/1502.01657, 2015.

\bibitem{hearn-13}
Mike Hearn.
\newblock {M}erge {A}voidance.
\newblock \url{https://medium.com/p/7f95a386692f}.
\newblock Accessed: 2016-03-01.

\bibitem{imwinkelried-et-al-15}
Edward Imwinkelried and Jason Luu.
\newblock {T}he {C}hallenge of {B}itcoin {P}seudo-{A}nonymity to {C}omputer
  {F}orensics.
\newblock 2015.

\bibitem{lischke-fabian-16}
Matthias Lischke and Benjamin Fabian.
\newblock {A}nalyzing the {B}itcoin {N}etwork: {T}he {F}irst {F}our {Y}ears.
\newblock {\em Future Internet}, 8(1):7, 2016.

\bibitem{maxwell-13}
Gregory Maxwell.
\newblock {C}oin{J}oin: {B}itcoin {P}rivacy for the {R}eal {W}orld.
\newblock \url{https://bitcointalk.org/index.php?topic=279249}.
\newblock Accessed: 2016-03-01.

\bibitem{maxwell-13-1}
Gregory Maxwell.
\newblock {C}oin{S}wap: {T}ransaction {G}raph {D}isjoint {T}rustless {T}rading.
\newblock \url{https://bitcointalk.org/index.php?topic=321228}.
\newblock Accessed: 2016-04-01.

\bibitem{meiklejohn-et-al-13}
Sarah Meiklejohn, Marjori Pomarole, Grant Jordan, Kirill Levchenko, Damon
  McCoy, Geoffrey~M. Voelker, and Stefan Savage.
\newblock {A} {F}istful of {B}itcoins: {C}haracterizing {P}ayments among {M}en
  with {N}o {N}ames.
\newblock In {\em Proceedings of the 2013 Internet Measurement Conference {IMC}
  2013, Barcelona, Spain, October 23--25, 2013}, pages 127--140, 2013.

\bibitem{monaco-15}
John Monaco.
\newblock {I}dentifying {B}itcoin {U}sers by {T}ransaction {B}ehavior.
\newblock 2015.

\bibitem{moeser-et-al-13}
Malte M{\"o}ser, Rainer B{\"o}hme, and Dominic Breuker.
\newblock {A}n {I}nquiry into {M}oney {L}aundering {T}ools in the {B}itcoin
  {E}cosystem.
\newblock In {\em Proceedings of the APWG eCrime Researchers Summit, ECRIME
  2013}, San Francisco, USA, 2013.

\bibitem{moeser-et-al-14}
Malte M{\"o}ser, Rainer B{\"o}hme, and Dominic Breuker.
\newblock {T}owards {R}isk {S}coring of {B}itcoin {T}ransactions.
\newblock In {\em Proceedings of the First Workshop on Bitcoin Research in
  Assocation with Financial Crypto 2014}, pages 1--16, Barbados, 2014.

\bibitem{nakamoto-08}
Satoshi Nakamoto.
\newblock {B}itcoin: A {P}eer-to-{P}eer {E}lectronic {C}ash {S}ystem.
\newblock 2008.

\bibitem{bip-75}
Justin Newton, Matt David, Aaron Voisine, and James MacWhyte.
\newblock {O}ut of {B}and {A}ddress {E}xchange using {P}ayment {P}rotocol
  {E}ncryption.
\newblock \url{https://github.com/bitcoin/bips/blob/master/bip-0075.mediawiki}.
\newblock Accessed: 2016-04-01.

\bibitem{nick-15}
Jonas Nick.
\newblock {D}ata-{D}riven {D}e-{A}nonymization in {B}itcoin.
\newblock Master's thesis, ETH Z{\"u}rich, 8 2015.

\bibitem{ober-et-al-13}
Micha Ober, Stefan Katzenbeisser, and Kay Hamacher.
\newblock {S}tructure and {A}nonymity of the {B}itcoin {T}ransaction {G}raph.
\newblock {\em Future Internet}, 5(2):237--250, 2013.

\bibitem{ortega-13}
Marc Ortega.
\newblock {T}he {B}itcoin {T}ransaction {G}raph---{A}nonymity.
\newblock Master's thesis, Universitat Oberta de Catalunya, 6 2013.

\bibitem{ranvier-13}
Justus Ranvier.
\newblock {R}eclaiming {F}inancial {P}rivacy with {HD} {W}allets.
\newblock
  \url{http://bitcoinism.blogspot.ie/2013/07/reclaiming-financial-privacy-with-hd.html}.
\newblock Accessed: 2016-03-01.

\bibitem{bip-47}
Justus Ranvier.
\newblock {R}eusable {P}ayment {C}odes for {H}ierarchical {D}eterministic
  {W}allets.
\newblock \url{https://github.com/bitcoin/bips/blob/master/bip-0047.mediawiki}.
\newblock Accessed: 2016-04-01.

\bibitem{reid-harrigan-13}
Fergal Reid and Martin Harrigan.
\newblock {A}n {A}nalysis of {A}nonymity in the {B}itcoin {S}ystem.
\newblock In Yaniv Altshuler, Yuval Elovici, Armin Cremers, Nadav Aharony, and
  Alex Pentland, editors, {\em Security and Privacy in Social Networks}, pages
  197--223. Springer New York, 2013.

\bibitem{ruffing-et-al-14}
Tim Ruffing, Pedro Moreno{-}Sanchez, and Aniket Kate.
\newblock {C}oin{S}huffle: {P}ractical {D}ecentralized {C}oin {M}ixing for
  {B}itcoin.
\newblock In {\em Proceedsing of the 19th European Symposium on Research in
  Computer Security, Wroclaw, Poland, September 7--11, 2014}, pages 345--364,
  2014.

\bibitem{shentu-yu-15}
QingChun ShenTu and Jianping Yu.
\newblock {R}esearch on {A}nonymization and {D}e-anonymization in the {B}itcoin
  {S}ystem.
\newblock {\em CoRR}, abs/1510.07782, 2015.

\bibitem{spagnuolo-et-al-13}
Michele Spagnuolo, Federico Maggi, and Stefano Zanero.
\newblock {B}it{I}odine: {E}xtracting {I}ntelligence from the {B}itcoin
  {N}etwork.
\newblock In {\em Proceedings of the 18th International Conference on Financial
  Cryptography and Data Security, FC 2014, Christ Church, Barbados, March 3--7,
  2014}, pages 457--468, 2014.

\bibitem{todd-14}
Peter Todd.
\newblock {S}tealth {A}ddresses.
\newblock
  \url{https://lists.linuxfoundation.org/pipermail/bitcoin-dev/2014-January/004020.html}.
\newblock Accessed: 2016-04-01.

\bibitem{valenta-rowan-15}
Luke Valenta and Brendan Rowan.
\newblock {B}lindcoin: {B}linded, {A}ccountable {M}ixes for {B}itcoin.
\newblock In {\em Proceedings of the 19th International Conference on Financial
  Cryptography and Data Security, FC 2015, San Juan, Puerto Rico, January 30,
  2015}, pages 112--126, 2015.

\bibitem{ziegeldorf-et-al-15}
Jan~Henrik Ziegeldorf, Fred Grossmann, Martin Henze, Nicolas Inden, and Klaus
  Wehrle.
\newblock {C}oin{P}arty: {S}ecure {M}ulti-{P}arty {M}ixing of {B}itcoins.
\newblock In {\em Proceedings of the 5th {ACM} Conference on Data and
  Application Security and Privacy, {CODASPY} 2015, San Antonio, TX, USA, March
  2--4, 2015}, pages 75--86, 2015.

\end{thebibliography}
